\documentclass[aps,prl,twocolumn,showpacs,superscriptaddress]{revtex4-1}
\usepackage{graphicx}
\usepackage{amsmath}
\usepackage{amssymb}
\usepackage{epstopdf}
\usepackage{amsfonts}
\usepackage{dcolumn}
\begin{document}
\title{
Abnormal Superfluid Fraction of Harmonically Trapped Few-Fermion Systems
}
\author{Yangqian Yan}
\affiliation{Department of Physics and Astronomy,
Washington State University,
  Pullman, Washington 99164-2814, USA}
\author{D. Blume}
\affiliation{Department of Physics and Astronomy,
Washington State University,
  Pullman, Washington 99164-2814, USA}
\date{\today}

\begin{abstract}
  Superfluidity is a fascinating phenomenon that, at the macroscopic scale, leads
  to dissipationless flow and the emergence of vortices. While these macroscopic
  manifestations of superfluidity are well described by theories that
  have their origin in Landau's two-fluid model, our microscopic understanding
  of superfluidity is far from complete. Using analytical and numerical
  \textit{ab initio} approaches, this paper determines the superfluid fraction and
  local superfluid density of small harmonically trapped two-component Fermi gases
  as a function of the interaction strength and temperature. At low
  temperature, we find that the
  superfluid fraction is, in certain regions of the parameter space, negative.
  This counterintuitive finding is traced back to the symmetry of the system's ground state wave function, which gives rise to a diverging quantum moment of inertia $I_{\text{q}}$. Analogous
  abnormal behavior of $I_{\text{q}}$ has been observed in even-odd nuclei at low
  temperature. Our predictions can be tested in modern cold atom experiments.
\end{abstract}
\maketitle

Superfluidity plays a crucial role in various areas of physics.
The core
of neutron stars is thought to be superfluid, giving
rise to modifications of the specific heat and rapid cooling~\cite{page00,astro11}.
In laboratory settings,
the superfluidity of bosonic liquid helium-4 below 2.17K and fermionic liquid
helium-3 below 3mK leads to dissipationless flow
and the formation of vortices~\cite{tilley1990superfluidity}.
More recently, superfluidity has been demonstrated 
in various dilute atomic Bose and Fermi gas experiments
\cite{cornell99,madison00,abo2001observation,zwierlein2005vortices}.

Over the past 20 years or so, non-classical
rotations in small doped bosonic helium-4 and molecular para-hydrogen
clusters have been, through
combined theoretical
and experimental studies~\cite{toennies04,grebenev98,grebenev00,tang02,kwon02,kwon05},
interpreted within the framework of
microscopic superfluidity.
Some elements of this framework date back to 1959 
when Migdal introduced a moments of inertia based method for the
study of superfluidity in finite-sized
nuclei~\cite{migdal59}. In nuclei, superfluidity is tied to the pairing  
of nucleons~\cite{dean03,broglia2013fifty}. 
As a consequence of pairing, the quantum moment of inertia of
even-even nuclei, i.e., nuclei with an even number of protons and an even
number of neutrons, tends to go to zero in the zero temperature limit while
that of even-odd nuclei tends to increase sharply as the 
temperature approaches zero~\cite{liu05}.

The present work investigates the superfluid fraction and local superfluid density
of small dilute atomic Fermi gases over a wide range of
interaction strengths.
In the low temperature region, we identify parameter combinations where the
quantum moment of inertia is abnormally large, i.e., larger than
the classical moment of inertia, implying a negative superfluid fraction. The
negative superfluid fraction is linked to the topology of the density matrix.
Specifically, the superfluid fraction takes on negative values in the low
temperature regime when one of the energetically low-lying eigen states
supports a Pauli vortex with finite circulation~\cite{vortexrev,dirac31,hirschfelder74} 
at the center of the trap.
Intuitively, this can be
understood as follows: In the absence of a 
low-energy eigen state with finite circulation, the superfluid few-fermion
gas ``does not respond'' to an infinitesimal rotation. This situation closely
resembles that for a superfluid few-boson gas. In the presence of a 
low-energy eigen state with finite circulation,
however, the superfluid few-fermion gas ``responds strongly'' to an
infinitesimal rotation, i.e., the infinitesimal rotation leads to a dynamical
instability. We find that the radial superfluid density is negative near the trap
center and positive near the edge of the cloud, indicating that the dynamical
instability develops at the vortex core. A related
instability
also exists for bosonic few-atom systems. However, since the instability for 
bosons does not occur for an infinitesimal rotation but when the rotation
frequency is comparable to the angular trapping frequency~\cite{dagnino09}, the superfluid 
fraction, which is defined 
in the limit of infinitesimal rotation~\cite{leggett70,baym69,leggett73,ceperley87}, is not affected
by the instability. We note that a negative superfluid fraction has also been 
predicted to exist for the Fulde-Ferrell-Larkin-Ovchinnikov 
 state of fermions loaded into an optical lattice~\cite{paananen09}.

We consider $N$ atoms 
of mass $m$
described by the Hamiltonian $H$ in a spherically symmetric harmonic trap.
The system Hamiltonian under a small rotation 
about the $z$-axis
can, in the rotating frame, be expressed as 
$H_{\text{rot}}=H-\Omega {L}_z$~\cite{tilley1990superfluidity}, 
where $\Omega$ denotes the angular rotating frequency 
and
$L_z$ 
the $z$-component of the angular momentum operator
$\mathbf{L}$.
The
superfluid fraction $n_s$ is defined as
$n_s=1-I_{\text{q}}/I_{\text{c}}$~\cite{leggett70,baym69,leggett73,ceperley87},
where the quantum moment of inertia $I_{\text{q}}$ is 
defined in terms of the response
to an infinitesimal rotation, 
\begin{equation}
I_{\text{q}}=
\frac{\partial \langle L_z \rangle_{\text{th}}}{\partial \Omega}\bigg|_{\Omega=0},
  \label{QuantumMomentofInertiaDefinition}
\end{equation}
and 
$\langle \cdot \rangle_{\text{th}}$
indicates 
the
thermal average.
The classical moment of inertia $I_{\text{c}}$ is defined through $I_{\text{c}}=\langle m\sum_{n} r_{n,\perp}^{2} \rangle_{\text{th}}$,
where  $r_{n,\perp}$ is the distance
of the
$n$th particle to the rotating axis~\footnote{
  The distance 
$r_{n,\perp}$ between $\vec{r}_n$ and the axis $\hat{z}$ can be
  expressed as $r_{n,\perp}=\mid \vec{r}_n \times  \hat{z}\mid$.
}.

In this paper, we work in the canonical ensemble and determine the superfluid
fraction of small trapped systems as a function of the temperature $T$ using
two different approaches. (i) We use the path integral Monte Carlo
(PIMC)
approach to sample the density matrix at temperature
$T$~\cite{ceperleyrev,boninsegni05,contact13}. 
The superfluid
fraction $n_s$ 
and local superfluid density $\rho_s$ 
are
then obtained using the area 
estimator~\cite{ceperley89,ceperley87,kwon06}. 
(ii) We employ a
microscopic approach~\cite{contact13}: For the systems considered, 
$\mathbf{L}^2$ and $L_z$ commute with
the Hamiltonian $H$, implying that the total orbital angular momentum quantum
number $L$ and the corresponding projection quantum number 
$M$
are good
quantum numbers. After some algebra, one finds
$I_{\text{q}} = \hbar^2 \langle M^2 \rangle_{\text{th}} /(k_B T)$, 
where 
the thermal average runs over the system at rest~\cite{butler55}.
To evaluate $I_{\text{q}}$, we 
calculate
a large portion of the quantum mechanical
energy
spectrum and thermally average the quantity $M^2$.
From the same set of calculations, we determine $r_{n,\perp}^2$ (and
correspondingly $I_{\text{c}}$) via the
generalized virial theorem~\cite{tan1,tan2,tan3},
which applies to systems with short-range 
interactions with $s$-wave
scattering length $a_s$ under 
spherically symmetric
harmonic confinement with angular trapping frequency $\omega$,
$3 \omega^2 \sum_n\langle m r_{n,\perp}^{2} \rangle_{\text{th}} = 
2\langle E + a_{s} (\partial E / \partial a_{s})/2 \rangle_{\text{th}}$.
Here, $E$ denotes the total energy. 

We first consider $N$ 
identical non-interacting bosons or fermions described by 
the Hamiltonian $H=H_{\text{ni}}$,
\begin{equation}\label{GeneralHamiltonian}
  H_{\text{ni}}=\sum_{j=1}^{N}\left(\frac{-\hbar^2}{2m}\nabla_j^2+\frac{1}{2}m
  \omega^2\mathbf{r}_{j}^2
\right),
\end{equation} 
where $\mathbf{r}_j$ denotes the position vector of the $j$th atom.
The partition function of the non-interacting $N$-particle system
can be obtained 
by symmetrizing 
(anti-symmetrizing)
the single-particle partition function 
for bosons (fermions)~\cite{krauth2006statistical},\footnote{The permutations
  of the $N$-particle system are realized using the function IntegerPartitions
in Mathematica.}.
Using the $N$-body partition function,
we calculate the
thermal 
averages for $I_{\text{c}}$ and $I_{\text{q}}$
as follows.
By the virial theorem
~\footnote{The virial theorem is a special case of the generalized virial theorem with $a_s=0$},\cite{QMintrodunctionbook},
the classical moment of inertia is 
equal to $2 \langle E \rangle_{\text{th}}/(3 \omega^2)$.
To calculate $I_{\text{q}}$, we
note that
$\langle M^2 \rangle_{\text{th}}=\langle (\sum_{n}M_n)^2 \rangle_{\text{th}}$ 
simplifies
to $\langle\sum_n M_{n}^2 \rangle_{\text{th}}$ 
since the cross terms cancel;
here, $M_{n}$ is the projection quantum number 
of the $n$th particle.

\begin{figure}
\centering
\includegraphics[angle=0,width=0.35\textwidth]{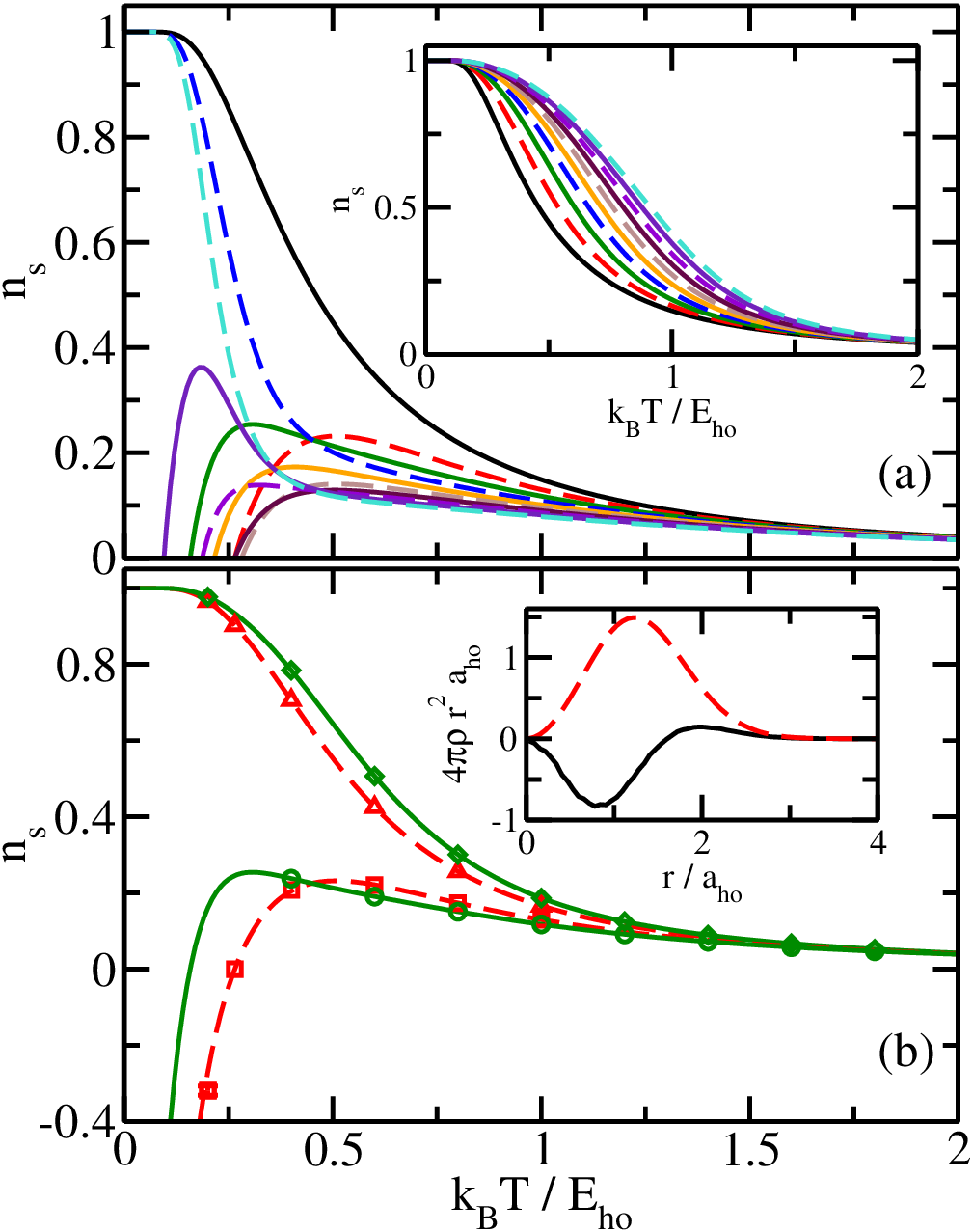}
\caption{(Color online)
  Superfluid fraction $n_s$ of 
the
non-interacting single-component gas as a function of $k_B T/E_{\text{ho}}$. (a) From top to bottom
  at $k_B T=E_{\text{ho}}$, the alternating solid and dashed lines show $n_s$ for the Fermi gas with $N=1-10$.
  Inset: From bottom to top, alternating solid and dashed lines show $n_s$ for the Bose gas with $N=1-10$.
  (b) The dashed and solid lines show $n_s$ for $N=2$ and $3$,
  respectively [these are the same data as those shown in
  Fig.~\ref{FigNonInteractingBosonsAndFermions}(a)]. For comparison, symbols show
  $n_s$ obtained using the PIMC approach. The error bars are smaller than the
  symbol size. Inset:
The dashed and solid lines 
  show the scaled radial total density and  scaled radial superfluid density for the two-fermion system.
 }\label{FigNonInteractingBosonsAndFermions}
\end{figure}

The inset of 
Fig.~\ref{FigNonInteractingBosonsAndFermions}(a) 
shows $n_s$ 
for $N=1-10$
non-interacting bosons.
For all 
$N$, 
 $n_s$ goes to 1 as
the 
temperature 
approaches zero. 
This is a direct
consequence of the fact that the ground state
has
$L=0$.
As the particle number increases,
 the superfluid region 
broadens.
The main panel of 
Fig.~\ref{FigNonInteractingBosonsAndFermions}(a) 
shows
$n_s$ for 
$N=1-10$ non-interacting fermions.
The curves have similar asymptotic
behavior at high temperature, yet they differ 
dramatically
at low temperature. 
The $N=1,4$ and $10$ 
curves 
increase monotonically with decreasing temperature and approach one
at $T=0$.
Due to the closed shell nature,
the ground state of these Fermi systems is, as that of the
Bose systems, non-degenerate
and has vanishing angular momentum.
The 
curves for the other $N$ values dive
down to negative infinity 
at
zero temperature.
The ground state of these open-shell systems is degenerate and contains finite angular momentum
states.
Figure~\ref{FigNonInteractingBosonsAndFermions}(b)
compares the analytical results 
(lines)
for $N=2$ and $3$ 
with those obtained by the
PIMC approach 
(symbols).
The 
excellent
agreement 
confirms 
the correctness of 
our
analytical
results
and demonstrates that
our PIMC
 simulations yield highly accurate results.
Although Stringari~\cite{stringari96}
determined $n_s$ for single-component Fermi
gases,
no
negative superfluid fraction 
was observed 
because the semi-classical 
treatment employed assumed
$k_B T \gg E_{\text{ho}}$, where $E_{\text{ho}}=\hbar \omega$.

To get a sense of the spatial distribution of the superfluid 
fraction, we calculate the radial superfluid density
$\rho_s(r)$~\footnote{
  $\rho_s(\mathbf{r})$ satisfies 
$I_{\text{c}}-I_{\text{q}}= m\int
\rho_s(\mathbf{r}) r_{\perp}^2 
d ^3 \mathbf{r}$,
where $r_{\perp}$ denotes the  distance to the 
$z$-axis.},\cite{kwon06}.
As an example, the solid line in the inset of
Fig.~\ref{FigNonInteractingBosonsAndFermions}(b) 
shows the scaled radial superfluid
density $\rho_s(r) r^2$ for the two-fermion system
at
$T=0.265 E_{\text{ho}}/k_B$.
For this temperature, we have $n_s=0$ (see main parts of
Fig.~\ref{FigNonInteractingBosonsAndFermions}).
The radial superfluid density is negative for small $r$
and positive for large $r$~\footnote{We note that the 
``direct'' integral 
$4 \pi \int_0^{\infty} \rho_s(r) r^2 dr$, 
i.e., the area under 
the curve, is not zero
since the radial superfluid density needs to be multiplied by an ``extra'' 
$r_{\perp}^2$ factor to yield the moment of inertia~\protect\cite{kwon06}.
}.
For comparison,
the dashed line 
shows the scaled radial total density.
For the non-interacting Fermi systems investigated, we find that the 
negative part of the radial superfluid density 
develops in the small $r$ region and then, with decreasing 
temperature, grows outward.

To interpret this behavior, we consider the $N=2$ case at $T=0$. In the absence of
rotation, the ground state has $L=1$ and the expectation value of $L_z$
averages to zero. The three-fold degenerate state splits under a small
rotation, with the $M=1$ state having the lowest energy; correspondingly, the
expectation value of $L_z$ is $\hbar$. Using these results to express $I_q$, 
see Eq.~(\ref{QuantumMomentofInertiaDefinition}), as a finite difference, we find
that $I_q$ scales as $\lim_{\Omega \to 0}\hbar \Omega^{-1}$ at $T=0$. This
analysis shows that the divergence of $I_q$ (and hence the negative value of
$n_s$) is due to the $M=1$ state, which contains a
vortex at the center of the trap with
circulation $1$. The inset of 
Fig.~\ref{FigNonInteractingBosonsAndFermions}(b)  shows that this is where the
radial superfluid density is negative, i.e., this is the region where the
dynamical instability develops.

Next, we consider two-component Fermi gases
consisting of $N_1$ spin-up
particles and $N-N_1$ spin-down particles
with short-range interspecies interactions.
As the $s$-wave scattering length is tuned from small negative values to 
infinity to small positive values, the system changes from
forming Cooper pairs to composite bosonic molecules~\cite{giorgini08}. 
In what follows we investigate how the change from ``fermionic''
(Cooper pairs) to ``bosonic'' (composite molecules) is reflected
in the superfluid properties of the system.
We consider the
Hamiltonian $H=H_{\text{int}}$,
\begin{equation}
  H_{\text{int}}=H_{\text{ni}}+
\sum_{j=1}^{N_1} \sum_{k=N_1+1}^{N}
V_{\text{tb}}(\mathbf{r}_{jk}),
  \label{HamiltonianWithInteraction}
\end{equation}
for
 two different interspecies
two-body potentials $V_{\text{tb}}$,
a 
regularized zero-range
pseudopotential $V_{\text{F}}$~\cite{Yang1957}
and a short-range Gaussian potential
$V_{\text{G}}$ with depth $U_0$ ($U_0<0$) and
range $r_0$,
$V_{\text{G}}(\mathbf{r}_{jk})=
U_{0} \exp 
[ -
\mathbf{r}_{jk}^2/(2r_{0}^2)
]$.
The depth and range are adjusted so that $V_{\text{G}}$
yields the desired $s$-wave scattering length $a_s$;
throughout, we consider potentials with $r_0 \ll a_{\text{ho}}$
[$a_{\text{ho}}=\sqrt{\hbar/(m \omega)}$]
that support at most one
free-space 
$s$-wave bound state.

\begin{figure}
\centering
\includegraphics[angle=0,width=0.35\textwidth]{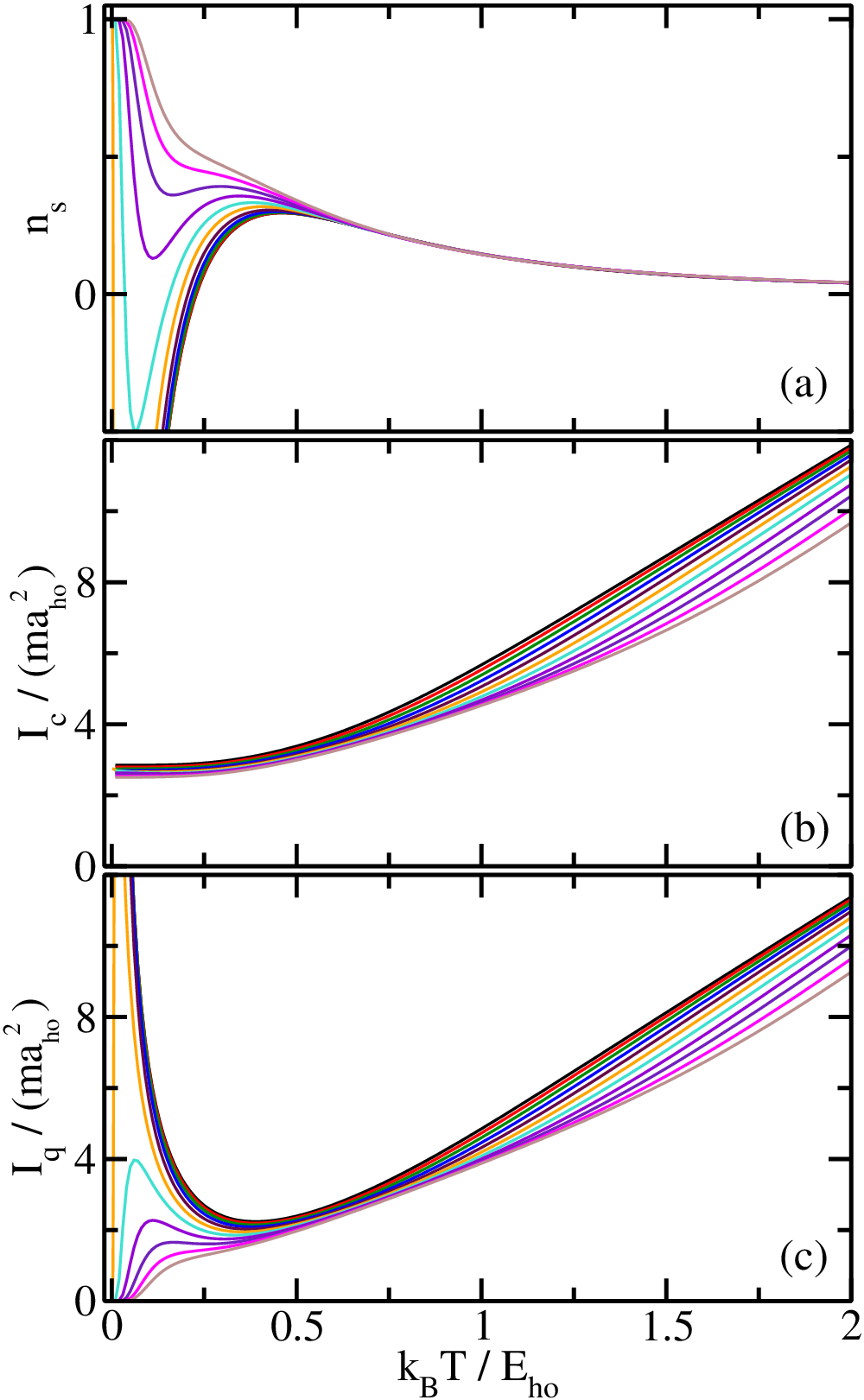}
\caption{(Color online)
  Properties of the interacting $(2,1)$ system as a function of $k_B T/E_{\text{ho}}$. (a) The lines from bottom to top show  $n_s$  for
 $a_{\text{ho}}/a_{s}=0,0.2,\dots,2$.
 (b)/(c) The lines from top to bottom show $I_{\text{c}}$
 and $I_{\text{q}}$, respectively,
 for
 $a_{\text{ho}}/a_{s}=0,0.2,\dots,2$.
 }\label{Fig21System}
\end{figure}

For the $(2,1)$ system
with zero-range interactions, 
we determine
a large portion of the
energy spectrum 
by solving the Lippman Schwinger equation
for arbitrary scattering length~\cite{duan07}.
This means that $n_s$ can be determined
within the microscopic approach over a wide temperature regime.
Figure~\ref{Fig21System}(b) 
shows 
the classical moment of inertia $I_{\text{c}}$ of the $(2,1)$ system 
as a function of the temperature $T$
for different $1/a_{s}$ ($a_s$ positive). 
$I_{\text{c}}$ decreases for fixed $T$
with
increasing $1/a_{s}$ and increases for fixed
$a_s$ with  increasing $T$. 
Figure~\ref{Fig21System}(c) 
shows the quantum moment of inertia $I_{\text{q}}$.
In the high temperature regime, $I_{\text{q}}$ and $I_{\text{c}}$
are nearly identical.
However, in the low temperature regime, notable differences exist.
For $1/a_s = 0$, $I_{\text{q}}$ 
diverges to positive infinity as $T \rightarrow 0$.
For $a_{\text{ho}}/a_s \approx 1$, in contrast, $I_{\text{q}}$ is zero
at $T=0$, increases 
sharply for $k_B T \lesssim 0.1 E_{\text{ho}}$, and then decreases 
for $k_B T \approx 0.1-0.5 E_{\text{ho}}$.
As $a_{\text{ho}}/a_s$ increases, the  local maximum moves to
larger temperatures and eventually disappears for $a_{\text{ho}}/a_s \approx 2$.
The dramatic change of $I_{\text{q}}$ at low $T$ on the positive scattering length 
side can be traced back to the symmetry change of the ground state wave function.
The lowest eigen state of the $(2,1)$ system has $L=1$ for 
$a_{\text{ho}}/a_s \lesssim 1$ and $L=0$ for 
$a_{\text{ho}}/a_s \gtrsim 1$.
Correspondingly, $I_{\text{q}}$ goes, in the zero $T$ limit, to $+\infty$ 
for
$a_{\text{ho}}/a_s \lesssim 1$ and to $0$ for 
$a_{\text{ho}}/a_s \gtrsim 1$.
The strong variation of $I_{\text{q}}$ near $a_{\text{ho}}/a_s \approx 1$
in the low  $T$  regime reflects the ``competing'' contributions of the
$L=0$ and $L=1$  states to the thermal average.

Combining $I_{\text{c}}$ and $I_{\text{q}}$ yields $n_s$ 
[see Fig.~\ref{Fig21System}(a)].
The $(2,1)$ systems with $a_{\text{ho}}/a_{s} \lesssim 1$ 
and $a_{\text{ho}}/a_{s} \gtrsim 1$ 
have a superfluid fraction that goes
to
negative infinity and one, respectively,
at zero temperature.
This can be viewed as a ``quantum phase transition like''
feature~\cite{duan07,conduit13}.
At $ k_B T = 0.2 E_{\text{ho}}$---a temperature
that might be achievable with current experimental set-ups~\cite{selim11,jochim13}---$n_s$ varies between 
$-0.14(1)$ and $0.54(1)$ 
for 
$a_{\text{ho}}/a_s =0$ to $2$.
For a given $a_s$, $n_s$ varies notably
over a small temperature regime. 
At high temperature ($k_B T \gtrsim 0.75 E_{\text{ho}}$), 
the effects of the 
$s$-wave interactions are less important
and $n_s$ is nearly independent of $a_s$.
The fact that $n_s$ is essentially independent of
$a_s$ for $k_B T \gtrsim 0.75 E_{\text{ho}}$ and
strongly dependent on 
$a_s$ for $k_B T \lesssim 0.4 E_{\text{ho}}$ might prove advantageous 
for qualitatively verifying the predicted behavior
experimentally.

\begin{figure}
\centering
\includegraphics[angle=0,width=0.35\textwidth]{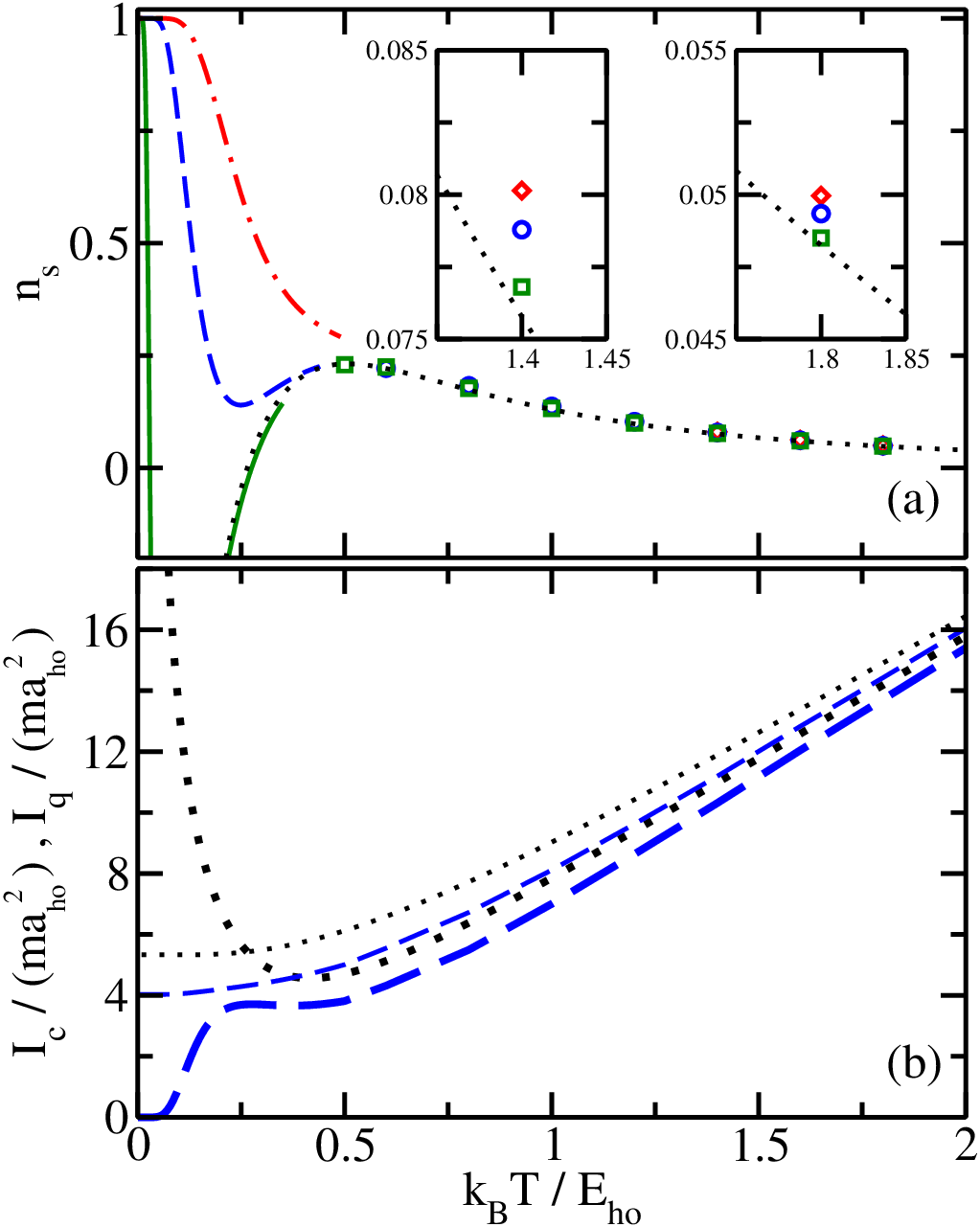}
\caption{(Color online)
  Properties of the $(2,2)$ system as a function of $k_BT/E_{\text{ho}}$. (a) The dotted, solid, dashed, and dash-dotted lines
  show  $n_s$ 
  for $a_{s}/a_{\text{ho}}=0,-0.2,-1,$ and $\infty$, respectively. 
  The squares, circles, and diamonds show $n_s$ 
  obtained by the PIMC approach for $a_{s}/a_{\text{ho}}=-0.2,-1,$ and $\infty$, respectively.
  The insets show blowups of the high-temperature region.
  (b) The thin dotted and dashed lines 
  show $I_{\text{c}}$ 
  for $a_{s}=0$ and $-a_{\text{ho}}$, respectively; 
  the thick dotted and dashed lines 
  show $I_{\text{q}}$ 
  for $a_{s}=0$ and $-a_{\text{ho}}$, respectively.
  The dashed curves are obtained by the microscopic approach 
(using
  $r_0=0.06a_{\text{ho}}$) for $ k_BT/E_{\text{ho}}\leq0.5$ and by the PIMC
  approach 
(using $r_0=0.1a_{\text{ho}}$) for $ k_BT/E_{\text{ho}}\geq0.6$.
 }\label{Fig22System}
\end{figure}

We now investigate what happens for a spin-balanced system.
Figure~\ref{Fig22System}(a)
shows $n_s$ for
the
 $(2,2)$ system 
with
$a_{s}/a_{\text{ho}}=0, -0.2, -1,$ and $\infty$. 
The ground state of the non-interacting $(2,2)$ system 
is nine-fold degenerate (one state has $L=0$, 
three states have $L=1$, and five
states have $L=2$).
The degeneracy of the ground state makes the 
quantum moment of inertia 
[see thick dotted line in Fig.~\ref{Fig22System}(b)]
diverge to plus infinity at $T=0$. 
The superfluid fraction, in turn, 
goes to minus infinity as $T \rightarrow 0$.
As the interactions are turned on, the degeneracy of the 
states with different $L$
is lifted, with the energy of the $L=0$ state lying below 
that of the
$L=1$ and $2$ states.
This implies that $I_{\text{q}}$ goes to zero at $T=0$
for $a_s \ne 0$
[for $a_s/a_{\text{ho}}=-1$, see 
  the thick dashed line
  in Fig.~\ref{Fig22System}(b)].
The behavior of the $(2,2)$ system is similar to that of the
$(2,1)$ system in that the zero temperature limit
of $n_s$ changes from minus infinity to one as the
scattering length is tuned. The transition, however, occurs at
different scattering lengths 
[$a_s=0$ for the $(2,2)$ system and $a_{\text{ho}}/a_s\approx1$
for the $(2,1)$ system].

\begin{figure}
\hspace*{0.1in}
\centering
\includegraphics[angle=0,width=0.35\textwidth]{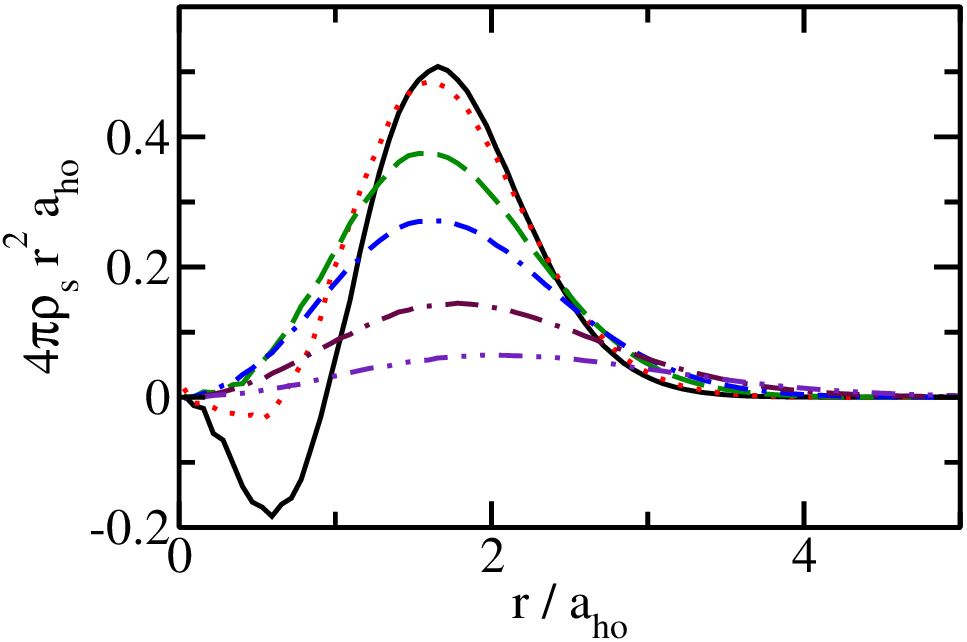}
\caption{(Color online)
  The solid, dotted, dashed, dash-dash-dotted, dash-dotted, and
  dash-dot-dotted lines show the scaled radial superfluid density $\rho_s r^2$
  for the 
$(2,2)$
system with scattering length 
$a_{s}=-0.2a_{\text{ho}}$ 
for $k_B T/E_{\text{ho}}=0.5,0.6,0.8,1,1.4,$ and $2$, respectively.
 }\label{FigRadialSuperfluidDensity}
\end{figure}

Figure~\ref{FigRadialSuperfluidDensity} 
shows 
the radial superfluid density for
the
$(2,2)$ system with $a_{s}=-0.2a_{\text{ho}}$
for various temperatures.
For the lowest temperature considered ($k_B T=0.5 E_{\text{ho}}$),
$n_s$ is equal to $0.230(3)$. Although $n_s$ is positive,
the radial superfluid density is
negative in the small $r$ region, reflecting the admixture of finite $L$
states to the density matrix.
As the temperature increases, the amplitude of the negative part of the 
radial superfluid density decreases and moves to smaller $r$.
When the radial superfluid density is positive everywhere, it
roughly has the same shape as the total radial density (not shown) but
with significantly decreased amplitude. This shows that the
superfluid
density is, in this regime, distributed roughly uniformly throughout the
cloud and not localized primarily near the center or edge of the cloud.
We find similar behavior for other scattering lengths.

In practice, thermal equilibrium can not be reached if the confinement is
spherically symmetric. We have checked that our results hold qualitatively for
anisotropic traps provided that $|\omega_x-\omega_y|\ll\omega_x+\omega_y$.
Moreover, the abnormal behavior of $n_s$ and $I_{\text{q}}$ is also found for finite
rotating frequencies, provided that $\hbar \Omega \ll E_{\text{ho}}$. Instead of
probing the response to a rotation of the trap, it might be possible to
simulate the rotation (and the resulting effective magnetic field) by
applying an effective gauge field~\cite{cooper10}.

To summarize, we determined the superfluid fraction and local superfluid
density of small harmonically trapped two-component Fermi gases as functions of the $s$-wave scattering length and temperature. At low temperature, the
quantum moment of inertia behaves, in certain regimes, abnormal, i.e., it is
larger than the classical moment of inertia, yielding a negative superfluid
fraction. The abnormal behavior arises if one or more of the low-lying
eigen states have a finite circulation, i.e., support a vortex. The relevant
temperature is roughly $\lesssim 0.5E_{\text{ho}}/k_B$. Our predictions
are unique to small systems, since such low temperatures can only be reached in
few-fermion systems~\cite{selim11,jochim13} and not in large Fermi gases.

{\em{Acknowledgement:}}
Support by the National
Science Foundation (NSF) through Grant No.
PHY-1205443
is gratefully acknowledged.
This work used the Extreme Science and Engineering
Discovery Environment (XSEDE), which is supported by
NSF Grant No. OCI-1053575, and the
WSU HPC.

\end{document}